\documentclass[journal]{vgtc}                     

\onlineid{0}

\vgtccategory{Research}

\vgtcpapertype{please specify}

\title{Analysis of the Causes of Car Accidents in the United States of America in 2023: Gauge People Understanding of Data Visualisation}

\author{%
  Hamoud Alhazmi,
  Marcelo Morales, Jiachen Jiang, Jinxin Zhou, and Jian Chen
}

\authorfooter{
  \item
    All authors are with Ohio State University.
    
  \item
  Hamoud Alhazmi. E-mail: alhazmi.8@osu.edu
    \item
    Marcelo Morales. E-mail: morales.374@osu.edu
    \item 
    Jiachen Jiang. E-mail: jiang.2880@osu.edu

    \item 
    Jinxin Zhou. E-mail: zhou.3820@osu.edu

    \item 
    Jian Chen. E-mail: chen.8028@osu.edu
  
}

\abstract{%
 This paper presents a comprehensive examination of interactive data visualization tools and their efficacy in the context of United States car accident data for the year 2023. We developed interactive heatmaps, histograms, and pie charts to enhance the understanding of accident severity distribution over time and location. Our research included the creation and distribution of an online survey, consisting of nine questions designed to test participants' comprehension of the presented data. Fifteen respondents were recruited to complete the survey, with the intent of assessing the effectiveness of both static and interactive versions of each visualization tool.
The results indicated that participants using interactive heatmaps showed a greater understanding of the data, as compared to those using histograms and pie charts. In contrast, no notable difference in comprehension was observed between users of static and interactive histograms. Unexpectedly, static pie charts were found to be slightly more effective than their interactive counterparts.
These findings suggest that while interactive visualizations can be powerful, their utility may vary depending on the type and complexity of the data presented. Future research is recommended to explore the influence of socioeconomic factors on the understanding of car accident data, potentially leading to more tailored and effective visualization strategies. This could provide deeper insights into the patterns and causes of car accidents, facilitating better-informed decision-making for stakeholders. 
  Visit our website to explore our interactive plots and engage directly with the data for a more comprehensive understanding of our findings. 
  \begin{itemize}
    \item URL: \url{https://cse-5544-datavis-project.vercel.app/}
\end{itemize}
}


\renewcommand{\manuscriptnotetxt}{}

\graphicspath{{figs/}{figures/}{pictures/}{images/}{./}} 

\usepackage{tabu}                      
\usepackage{booktabs}                  
\usepackage{lipsum}                    
\usepackage{mwe}                       

\usepackage{mathptmx}                  

\begin{document}


\firstsection{Introduction and Motivation}
\maketitle

Visualizing complex data remains crucial for several sectors, including public safety, urban planning, and transportation management. Among the many types of data, one that remains very important is the distribution of the severity of the accident at different times of the day and in various geographic areas. This understanding can offer great help to the decision process and therefore in the implementation of interventions and safety measures. Many static and interactive plots are used equally in the context of data visualization, each with its advantages and disadvantages. This research examines these different visualization techniques and compares them to determine which method best provides the accident distribution information.

There have been recent advances in interactive visualization technology, of which much has improved in terms of how a user interacts with data toward dynamically exploring complex information with understanding. An interactive plot allows the user to directly interact with the data. This involves dragging views, drilling down and filtering out information. These actions enable the user to gain deep insight from the data. This prompts a few critical inquiries. Firstly, do these interactive features continue to hold value? More essentially, does the interactivity truly augment our comprehension of the distribution of accident severity, particularly when juxtaposed with a conventional static plot?

This research is motivated by the need to:

\begin{itemize}
    \item Evaluate the Effectiveness of Interactive Visualization: By directly comparing interactive and static visualizations, we aim to assess which method better supports comprehension of complex datasets.
    
    \item Identify Optimal Plot Types: This study aims to determine the most effective form of interactive plots-HeatMap, Histogram, and PieChart-to communicate data on the distribution of accident severity. The selection of the type of plot can substantially affect the efficiency of information transmission and comprehension.
    
    \item Guide Future Visualization Tools: The results of this research will offer crucial insights that can steer the creation of upcoming visualization instruments, guaranteeing their ease of use and efficacy in conveying vital data points.
\end{itemize}

By providing a comprehensive view of how visualization techniques can be leveraged, this study offers practical insights into improving public safety through improved data interpretation and decision processes. These findings can be directly applied to the development of user-friendly visualization tools.

\section{Research Questions}

The primary objective of this study is to examine the effectiveness of different visualization techniques in helping people understand the distribution of the severity of the accident over time and locations. To achieve this, we have formulated the following specific research questions:
\begin{itemize}
    \item Is an interactive plot more effective than a static plot in helping people understand the distribution of accident severity over time and locations?
    \item Among interactive graphs such as HeatMap, Histogram, and PieChart, which is the most effective to aid in understanding the distribution of the severity of the accident?
\end{itemize}

The first question aims to determine whether the dynamic display of interactive plots would yield better data understanding than that of static plots with a fixed format. The second question will establish which type of interactive visualization, HeatMap, Histogram, or PieChart, could best be used to display complex spatial and temporal data. It should be used to understand how the structure and presentation styles of the different plots facilitate user comprehension and interaction.

\section{Preliminary Hypotheses}

Based on the research questions outlined, we propose the following hypotheses to guide our experimental investigation:
\begin{itemize}
    \item Interactive plots will be more effective than static plots in helping people understand the distribution of accident severity over time and locations.
    
    \item Among the types of interactive plots, HeatMap will prove to be the most effective in helping people understand the distribution of severity of accidents.
\end{itemize}

Interactive plots allow users to use functionalities to engage deeply in the data, e.g. zooming, filtering, selecting specific parameters, etc. This capability improves comprehension and information retention since the facility enables learners to learn at their own pace and style.

By nature, HeatMap offers an intuitive geographical context and is thus very suitable for visualizing spatial data. It will be hypothesized that the familiar format of maps, combined with interactive elements, offers faster and more precise information on regional and temporal trends of accident severity.

\begin{figure}[tb]
  \centering 
  \includegraphics[width=\columnwidth, alt={A line graph showing paper counts between 0 and 160 from 1990 to 2016 for 9 venues.}]{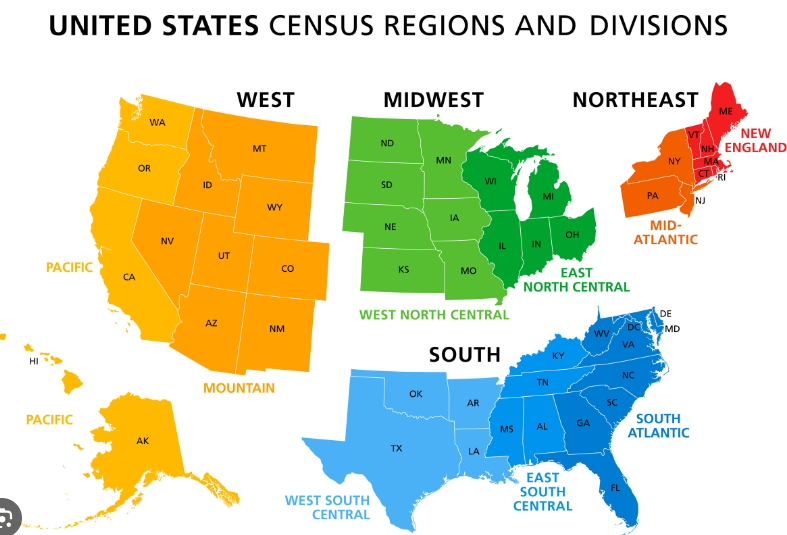}
  \caption{%
  	A visualization of  car accidents by regions to compare the severity per region.%
  }
  \label{fig:vis_papers}
\end{figure}

\section{Related Work}

In recent years, there has been a notable increase in the incidence of fatal car accidents in all states. In the United States, there were 42,795 fatal vehicle crashes in 2022 compared to 36,835 fatal car crashes in 2018, marking an increase of more than 16\% \cite{report2022, report2023}. The dataset we intend to utilize, consisting of more than 2.25 million traffic accident reports collected for the contiguous United States over a three-year period, represents the first nationwide dataset of this magnitude. 

Related work focuses on the prediction of real-time traffic accidents \cite{moosavi2019accident, moosavi2019countrywide}. However, our approach diverges as we aim to present the temporal distribution of car accidents by showcasing crashes by time of day. Furthermore, we intend to present various visual representations to effectively tackle the issue. This includes utilizing a histogram to depict the severity of the crash. We also plan to use a histogram showing the distribution of accidents over time and severity levels. In addition, a HeatMap will be used to indicate accident locations. Lastly, a gradient map will be used to illustrate the timing of the most severe accidents, among others. 

This project addresses the relationship between crash statistics in the United States of America and how people perceive car accidents.

\section{Evaluation Consideration }

There are numerous visualization and analysis techniques that can be considered to explore the US-Accidents dataset. However, we consider the utilization of environmental factors, temporal analysis, comparative analysis, and spatial analysis. The aforementioned analysis will provide us with a complete picture of the problem. 

Additionally, our intention is to present data on car accidents in each contiguous state, followed by their classification according to regions, to compare the severity within each region as illustrated in Figure 1. On the other hand, to improve the performance, we think of creating multiple interactive plots and combining them using Tableau. We will explore evaluation methods similar to \cite{kahn2015national,setiawan2019visualization, shaadan2021road, johansson2004carsim}.

\section{Experimental Design}
The experiment design aims to systematically compare the effectiveness of interactive and static visualizations to help to understand the distribution of the severity of the accident over time and locations. The design is structured around two main components: figure design and survey design.

\subsection{Hypothesis 1: Interactive vs. Static Maps}
To effectively test the first hypothesis, that interactive maps are more effective than static plots in helping people understand the distribution of the severity of accidents over time and locations—the experiment design will include a specific focus on comparing interactive and static maps. Here is a step-by-step breakdown of how this will be implemented:

The interactive map will allow users to select different time segments (morning, afternoon, evening, night) and regions (West, Midwest, South, Northeast). This can be done through dropdown menus. Interactive elements such as zoom, hover details (showing accident severity data upon hovering over a region), and dynamic filtering (to isolate specific data points based on user selection) will be included. Color gradients will be used to represent varying levels of accident severity, with tooltips providing quantitative details.

For each of the timeslices, we will create a sequence of static maps to distribute the severity of the accidents over the four areas of interest. Each map will use an identical color gradient to the interactive one, only for consistency. However, none of these maps will have interactive capabilities. The maps will be presented in sequence or grid format, allowing the participant to view differences between time and region of the forecast without interaction with a view.

\subsection{Hypothesis 2: Different Types of Plots}
To test the second hypothesis -that graphs are the most effective type of interactive graph to help people understand the distribution of accident severity - our experimental design will compare interactive HeatMap, Histogram, and PieChart. The detailed experimental setup is below for evaluating the effectiveness of each visualization type.

HeatMap will provide geographical context with zoom, filter, and hover-over details (such as specific accident severity data). Users can select different times and regions to view specific data, enhancing spatial understanding.

The histogram will be designed to show the totals of the severity of the accident for different regions and times. Users can click on specific bars to drill down into more detailed temporal or regional data. Users can also use the filter to select the range of total severity numbers of different states.

PieChart will depict accident severity data as a function of time of day, with different pie segments representing different states. Clicking on a segment can reveal more detailed data, and users can toggle between different layers of information.

\subsection{Survey Design}
The study used a methodological approach involving 15 participants, segmented into three groups of five. Each group was tasked with evaluating a different data visualization format: one for heatmaps, another for histograms, and the third for pie charts. This approach facilitated a targeted comparison of user interaction and comprehension between the various visualization methods. After interacting with the maps, participants will answer a series of survey questions that assess their comprehension and ability to use the information presented effectively. 

These will include both direct questions about specific data points and more general questions about trends and patterns. Tasks include identifying the time of day or region with the highest severity of the accident or comparing severity between different regions or times.

The responses will be evaluated for correctness. The time each participant takes to complete the tasks will be recorded. This will help determine whether interactive features enhance quicker comprehension compared to static visuals.

The data collected will be analyzed using statistical methods to compare performance metrics (accuracy and response time) between all groups.
Effectiveness will be measured based on the speed and accuracy of responses, providing quantitative evidence on whether interactive maps lead to better or faster understanding compared to static maps.

\section{Implementation Method}
In this section, we detail the specific methods and technologies employed to support our hypothesis. The strength of our execution approach is crucial to guaranteeing the dependability and reproducibility of our findings. 

\subsection{Interactive Plots}
Interactive plots are particularly effective in multivariate data analysis, where dynamic adjustment of variables can reveal hidden relationships and influences between data points. By incorporating interactive graphs into our analysis, we provided a user-friendly interface using Tableau that allowed reviewers and readers to explore data dependencies and correlations at their discretion. 

The website provides an intuitive and user-friendly interface, allowing visitors to effortlessly navigate through various interactive plots and gain insights without needing extensive technical knowledge. With its clean layout and responsive design, the website ensures that users of all devices can access and interact with the plots seamlessly, enhancing the user experience across different platforms.

Each interactive plot on the website is accompanied by concise, informative captions and contextual help, guiding users through the data exploration process without overwhelming them. There are three interactive plots on the website: HeatMap, Histogram, and PieChart as shown in Figures Figure 2, Figure 3, and Figure 4. The accidents and their severity can be viewed by users by navigating through them and choosing the region and time.

\begin{figure}[tb]
  \centering 
  \includegraphics[width=\columnwidth, alt={A line graph showing paper counts between 0 and 160 from 1990 to 2016 for 9 venues.}]{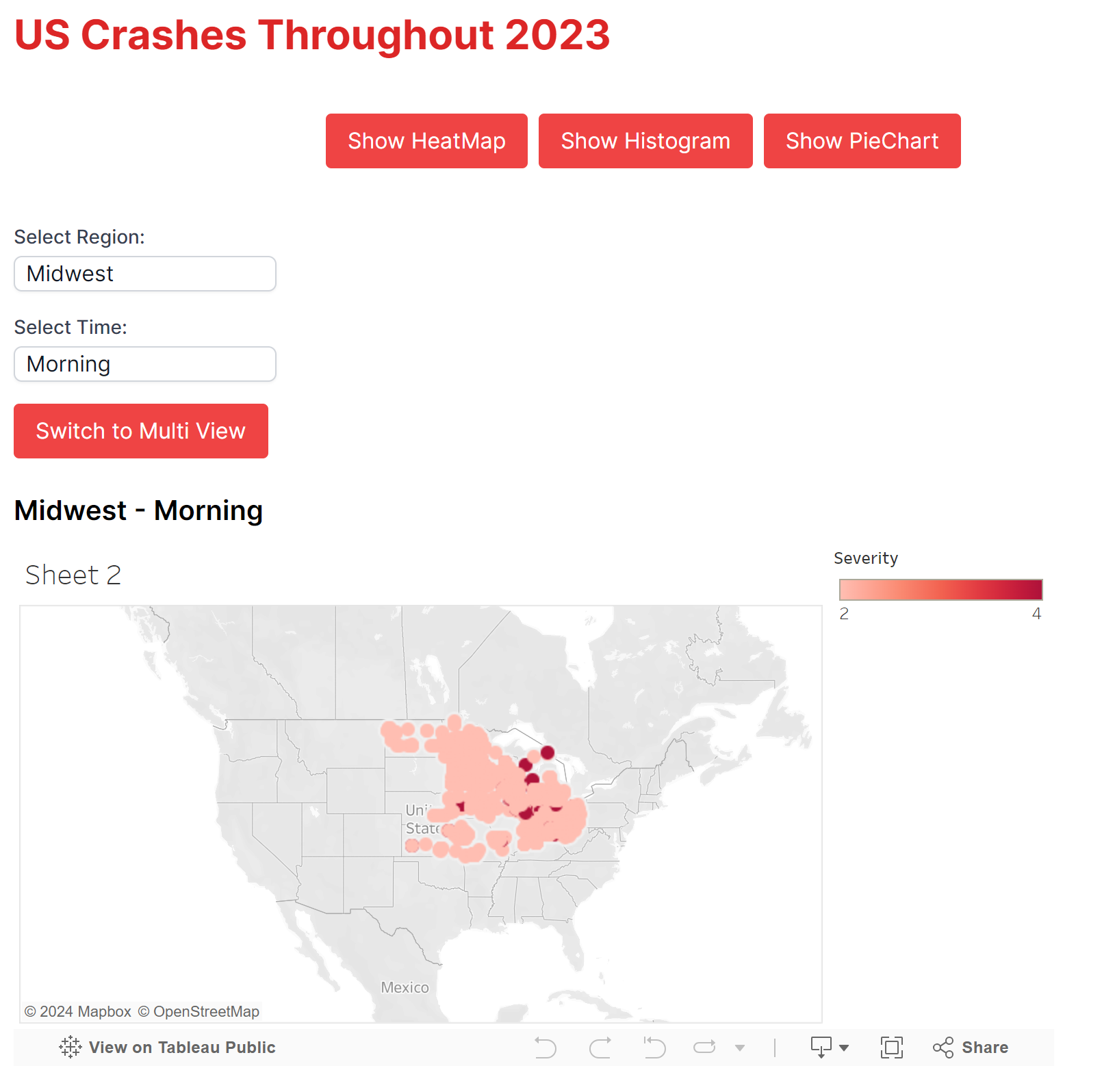}
  \caption{%
  	Interactive visualization of HeatMap provided in our website.%
  }
  \label{fig:vis_papers}
\end{figure}

\begin{figure}[tb]
  \centering 
  \includegraphics[width=\columnwidth, alt={A line graph showing paper counts between 0 and 160 from 1990 to 2016 for 9 venues.}]{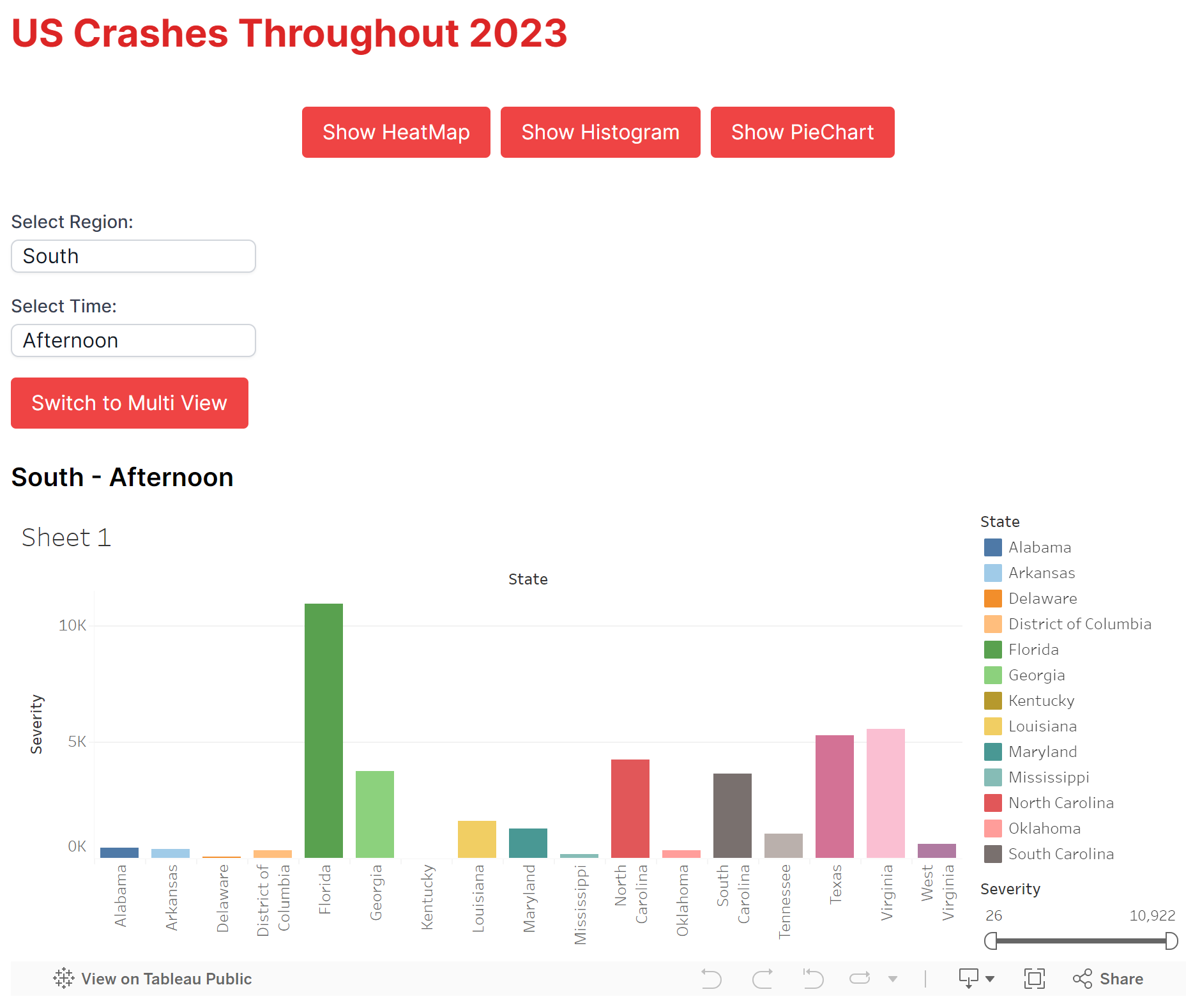}
  \caption{%
  	Interactive visualization of Histogram provided in our website.%
  }
  \label{fig:vis_papers}
\end{figure}

\begin{figure}[tb]
  \centering 
  \includegraphics[width=\columnwidth, alt={A line graph showing paper counts between 0 and 160 from 1990 to 2016 for 9 venues.}]{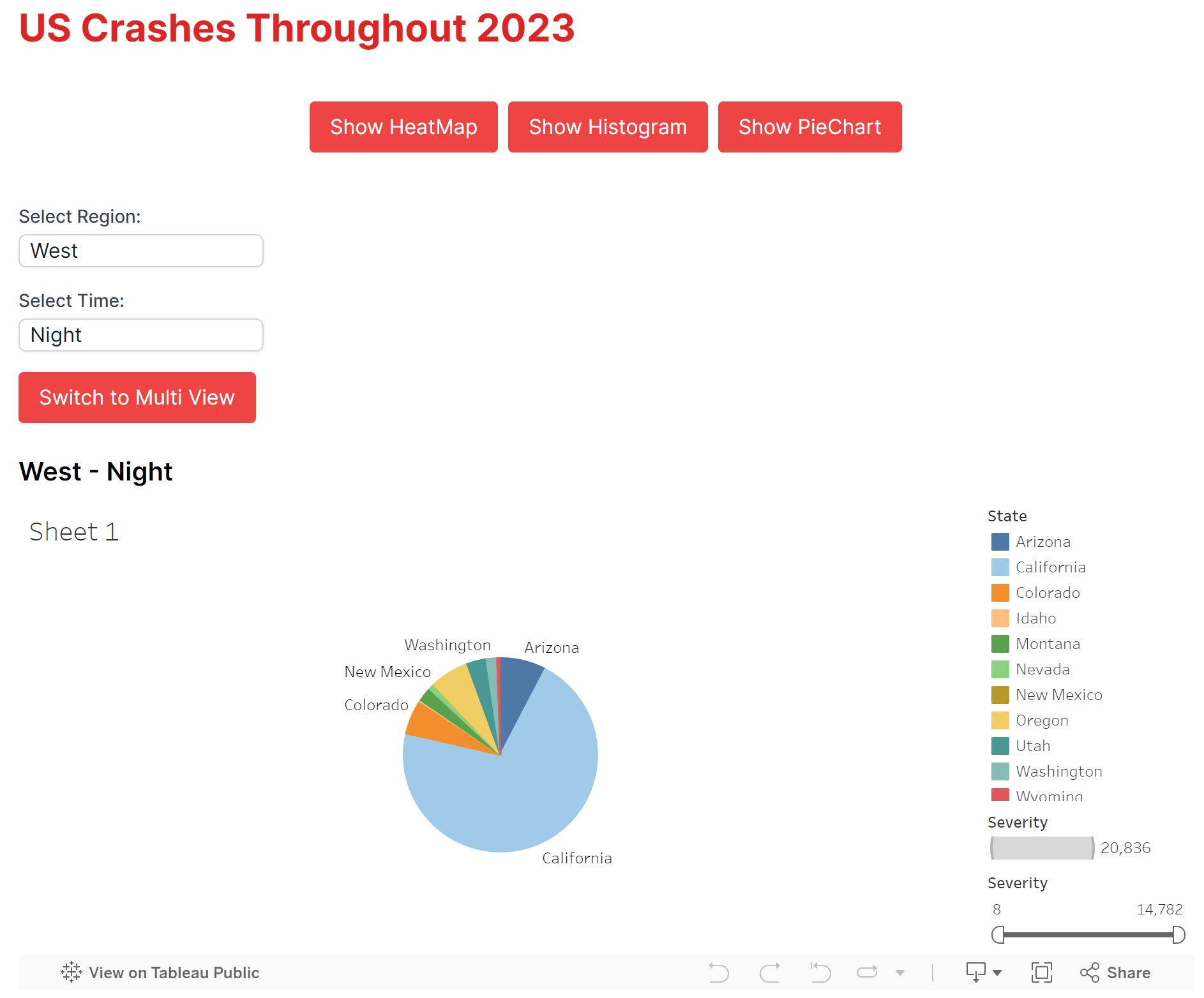}
  \caption{%
  	Interactive visualization of PieChart provided in our website.%
  }
  \label{fig:vis_papers}
\end{figure}

Visit our website to explore our interactive plots and engage directly with the data for a more comprehensive understanding of our findings. 

\begin{itemize}
    \item URL: \url{https://cse-5544-datavis-project.vercel.app/}
    
    \item The source code is available at: \url{https://github.com/mm587517/cse-5544-datavis-project}
\end{itemize}

\subsection{Survey Study}
The surveys were conducted using Google Forms, chosen for its accessibility and ease of use. Google Forms facilitated the rapid deployment of our survey, allowing us to collect responses in real time and immediately begin analyzing trends and patterns in the data. We developed the survey based on the hypothesis. In particular, the survey includes nine questions, as depicted in Figure 5. 

To assess understanding, the survey included multiple choice questions that tested participants' knowledge of key concepts, allowing us to measure their grasp of the information presented. In addition, the survey used scenario-based questions to simulate real-life situations, helping us to assess how well participants understood and could apply information in practical contexts. Our research team crafted a series of distinct surveys, three in total, each aimed at measuring the level of understanding on specific topics (i.e., HeatMap, Histogram, and PieChart) among diverse participant groups.

\begin{figure}[tb]
  \centering 
  \includegraphics[width=\columnwidth, alt={A line graph showing paper counts between 0 and 160 from 1990 to 2016 for 9 venues.}]{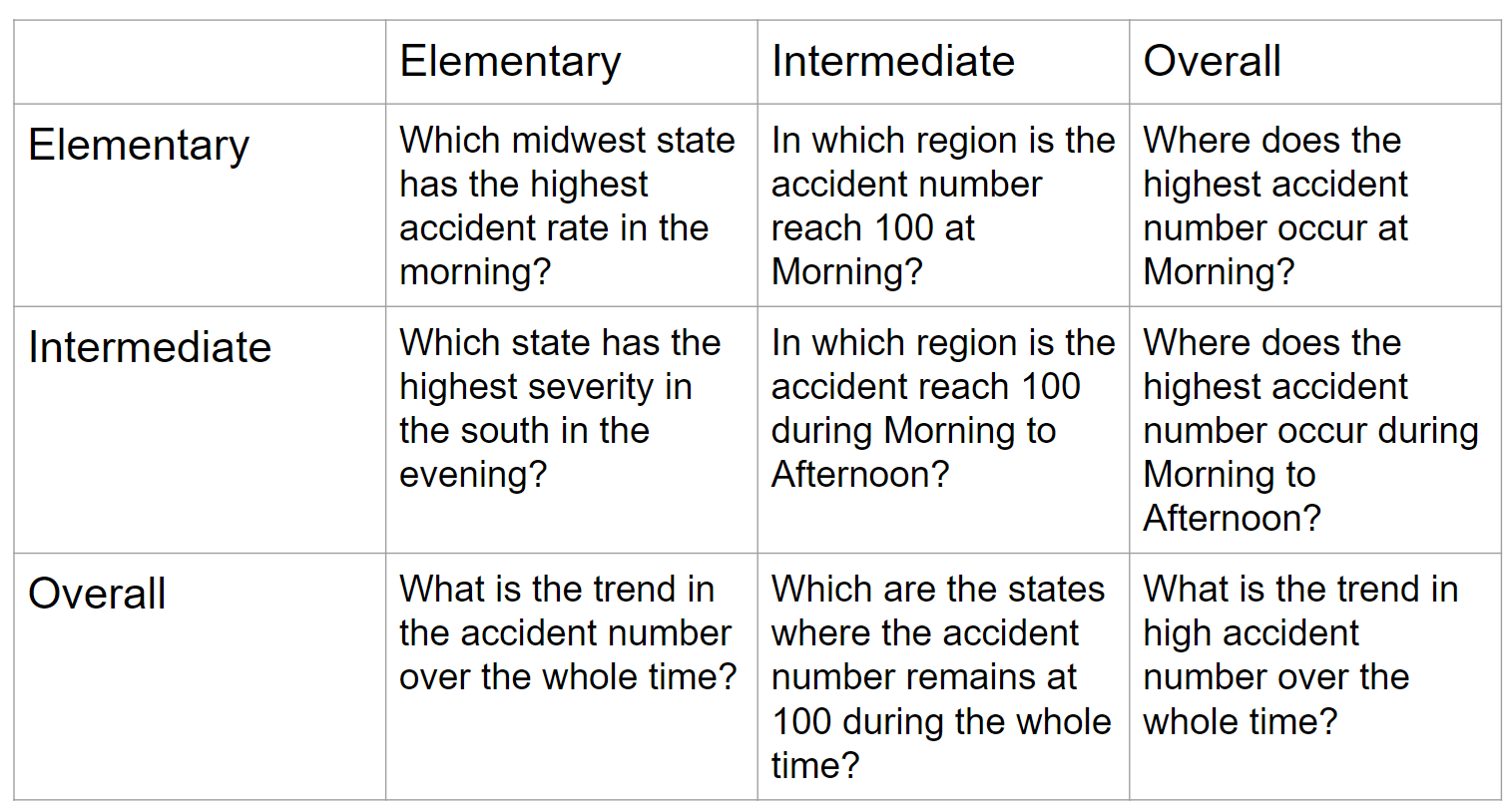}
  \caption{%
  	Survey questions to gauge participants' perceptions.%
  }
  \label{fig:vis_papers}
\end{figure}

\section{Results and Analysis}

Three distinct surveys were designed to evaluate the effectiveness of different data visualization tools: heatmaps, histograms, and pie charts. Each survey was administered to a separate group of five participants, specifically selected to assess one type of visualization tool. 

Consequently, a group of five participants evaluated the heatmap, another group assessed the histogram, and the final group analyzed the pie chart. This methodological approach ensured a focused and direct comparison of user interaction and comprehension across different visualization formats.

\subsection{Static vs. Interactive Charts}

\subsubsection{HeatMap}

According to our analysis in comparing the effectiveness of interactive versus static heatmaps, survey respondents have been observed to demonstrate significantly higher proficiency with interactive models. More than 75\% of the participants accurately answered questions when using interactive heatmaps, indicating a substantial improvement in understanding and participation, as shown in Figure 6.

This increased performance can be attributed to the dynamic nature of interactive heatmaps, which allows users to manipulate the data visually, explore different layers of information, and gain insights through active learning. Such interactive features are notably absent in static heatmaps, where data presentation is fixed, limiting the viewer's ability to delve deeper into the specifics of the data. This clear disparity underscores the importance of interactive elements in data visualization tools for improved comprehension and analytical capabilities.

\subsubsection{Histogram}

Interestingly, in Figure 6, when the survey respondents were presented with static and interactive histograms to analyze and answer the questions, the results did not show significant differences in their performance. This unexpected outcome suggests that the addition of interactive features in histograms may not universally improve user understanding or engagement, unlike other forms of data visualization such as heatmaps. 

The similarity in the results between static and interactive histograms could indicate that the simplicity and straightforward presentation of histograms, which typically display distributions and frequencies, may be sufficient for effective data comprehension. This finding challenges the assumption that interactivity is always beneficial in data visualization tools and highlights the need to consider the specific data type and user context when designing these tools.

\subsubsection{PieChart}

Surprisingly, the results of the survey in Figure 6 revealed that respondents performed slightly better when using static pie charts compared to their interactive counterparts. This counterintuitive finding suggests that the simplicity and directness of static pie charts may facilitate faster comprehension and decision-making. Static pie charts provide a clear and unchanging representation of the proportions of data, which might reduce the cognitive load and streamline the process of data interpretation. 

In contrast, interactive features in pie charts, while potentially enriching the user experience with additional data layers or insights upon interaction, could introduce complexities that momentarily distract or confuse users. This suggests that the benefits of interactivity in data visualization are not universally applicable and may depend on the context and nature of the data presented.

\begin{figure}[tb]
  \centering 
  \includegraphics[width=\columnwidth, alt={A line graph showing paper counts between 0 and 160 from 1990 to 2016 for 9 venues.}]{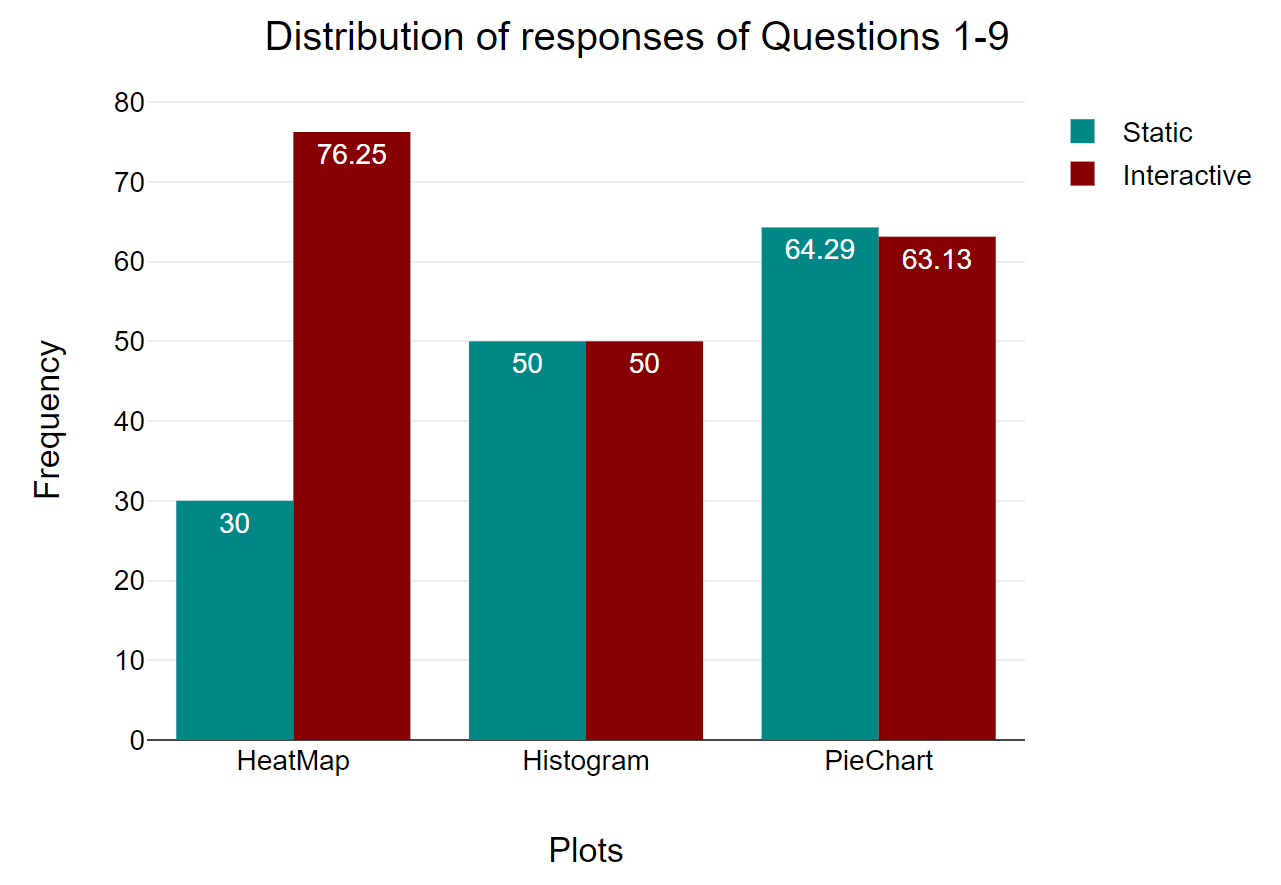}
  \caption{%
  	The distribution of responses of Questions 1-9 (HeatMap, Histogram, and PieChart).%
  }
  \label{fig:vis_papers}
\end{figure}

\subsection{Optimizing Data Insights with Comparison Charts}

As suggested in the preliminary hypothesis section, heat maps are hypothesized to enhance the comprehension of the accident severity distribution more effectively than histograms and pie charts, as illustrated in Figure 7. Quantitatively, the median values extracted from the surveys underscore this assertion: the median severity rating observed in the heatmap is 72.5, significantly higher than the 36.5 recorded for the histogram, and substantially surpassing the 21.5 noted in the pie chart. This data supports the hypothesis that heatmaps provide a superior visualization mechanism to detect complex data distributions.

\begin{figure}[tb]
  \centering 
  \includegraphics[width=\columnwidth, alt={A line graph showing paper counts between 0 and 160 from 1990 to 2016 for 9 venues.}]{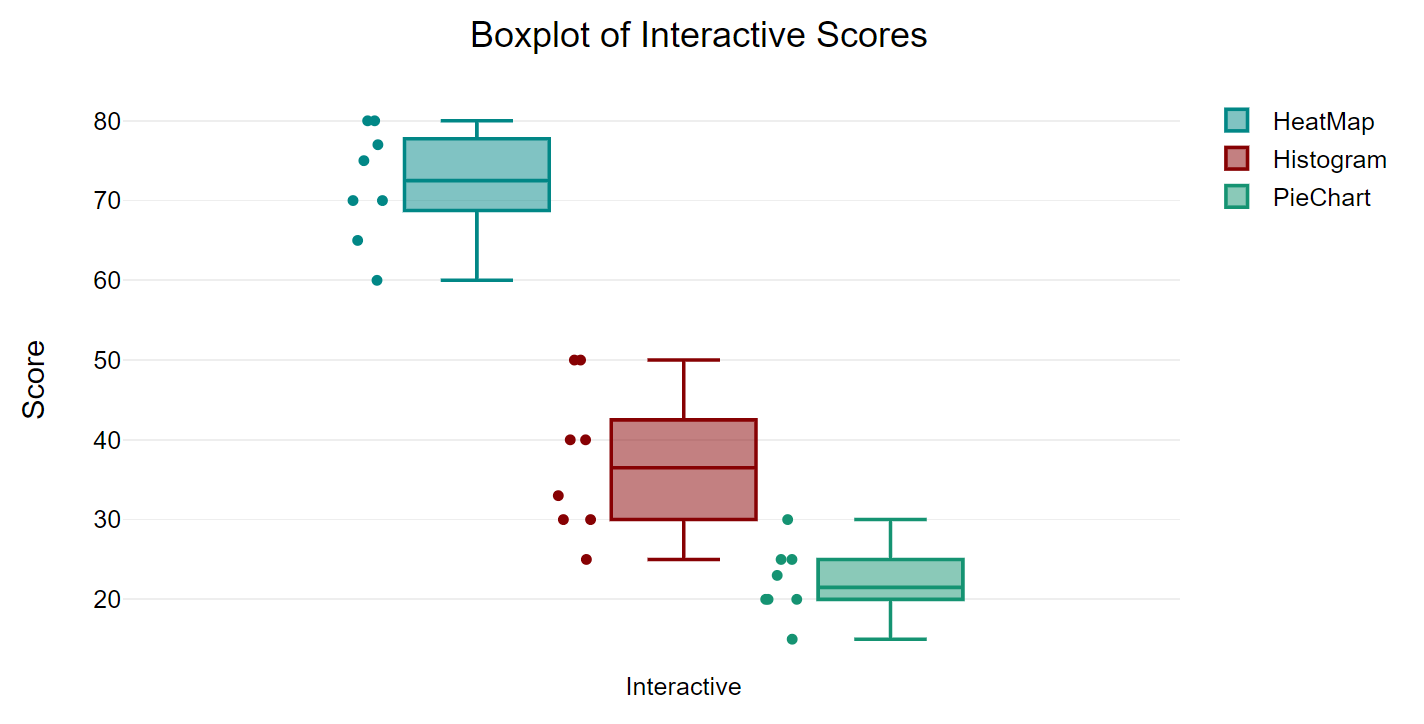}
  \caption{%
  	Comparative Analysis of Visualization Tool Efficacy: A Boxplot of Interactive Heatmap, Histogram, and Pie Chart Scores.%
  }
  \label{fig:vis_papers}
\end{figure}

\section{Conclusion and Future Work}

In the current study, we developed a series of interactive visual representations - namely, heat maps, histograms, and pie charts - to analyze vehicular accidents across the United States for the year 2023. These interactive visual tools have proven to be instrumental in improving our understanding of the complexities surrounding the distribution of the severity of the accident in temporal and spatial dimensions. To evaluate the effectiveness of these visual tools, we constructed an online survey consisting of nine meticulously crafted questions aimed at gauging the participants' grasp of the severity distribution data. Fifteen respondents were selected to participate in this survey.

The findings of the survey are illuminating. Interactive heat maps yielded the most favorable results, suggesting a higher efficacy in conveying complex data compared to histograms and pie charts. Interestingly, when comparing the histograms, no significant differences in performance were discerned between their interactive and static forms. In contrast, static pie charts unexpectedly surpassed their interactive variants in terms of user comprehension.

Taking these insights into account, future research might benefit from examining socioeconomic variables, such as income, age, marital status, occupation, and educational background, to enrich the analysis. An in-depth understanding of these factors may prove pivotal in unraveling the multifaceted nature of the severity of traffic accidents and its determinants.


\bibliographystyle{abbrv-doi-hyperref}
\bibliography{CSE5544_Project}

\end{document}